# Simulating Organogenesis in COMSOL: Comparison Of Methods For Simulating Branching Morphogenesis

Lucas D. Wittwer[1], Michael Peters[1,2], Sebastian Aland[3], Dagmar Iber[*,1,2]
[1]D-BSSE, ETH Zürich, Zurich, Switzerland
[2]Swiss Institute of Bioinformatics (SIB), Switzerland,
[3]Faculty of Informatics/Mathematics, University of Applied Sciences, Dresden, Germany
[*]Corresponding author: D-BSSE, ETH Zürich, Mattenstrasse 26, 4058, Switzerland, dagmar.iber@bsse.ethz.ch

**Abstract:** During organogenesis tissue grows and deforms. The growth processes are controlled by diffusible proteins, so-called morphogens. Many different patterning mechanisms have been proposed. The stereotypic branching program during lung development can be recapitulated by a receptor-ligand based Turing model [1]. Our group has previously used the Arbitrary Lagrangian-Eulerian (ALE) framework for solving the receptor-ligand Turing model on growing lung domains [2]–[4]. However, complex mesh deformations which occur during lung growth severely limit the number of branch generations that can be simulated. A new Phase-Field implementation avoids mesh deformations by considering the surface of the modelling domains as interfaces between phases, and by coupling the reaction-diffusion framework to these surfaces [5]. In this paper, we present a rigorous comparison between the Phase-Field approach and the ALE-based simulation.

## Introduction

During mouse lung development, thousands of branches form in a highly stereotypic manner to maximise the surface-volume ratio for gas exchange [6], [7]. The final developed organ consists of two lungs: the left lung consists of only one lung lobe, while the right lung contains four lobes. Each lobe consists of branched bronchioles made of epithelial tissue and surrounding connective mesenchymal tissue.

Based on wet-lab experiments, Metzger et al. [6] identified three local branching modes: (a) domain branching, where daughter branches appear along the main stalk, (b) planar bifurcation, where the tips split in the same plane several times and (c) orthogonal bifurcation, where the bifurcation is rotated by approximately 90 degrees compared to the last bifurcation event [6]. Thus, the first two modes fill the 2D space and only the last mode, the orthogonal bifurcation, fills the 3D volume. While those different modes are executed concurrently in different parts of the developing lung, the transition from one mode to another is restricted. A fourth mode known as trifurcation (d) has been observed in the embryonic lung, but it is much more common in the ureteric bud of the kidney [7].

Several mechanisms have been proposed to explain the branching dynamics in the embryonic lung, e.g. based on fractals, mechanical properties of the surrounding tissues or signalling models [8]. However, only a receptor-ligand based Turing mechanism can recapitulate the experimentally observed branching pattern [9].

Turing mechanisms –postulated by Alan Turing (1952) [10]– are reaction-diffusion systems which can produce stable, non-uniform, and symmetry breaking patterns like stripes or spots. Such a system contains two chemical species, so called morphogens, diffusing with different speeds and interacting in a non-linear manner. Many other patterning phenomena in biology have been accounted to Turing Patterns, e.g. fur patterns [11], [12]. Even though chemical reactions producing Turing Patterns have been found in the meantime [13], [14], a definitive proof in a biological system is still outstanding.

The receptor-ligand based Turing mechanism in the lung models the observed interaction between the ligand protein FGF10 denoted as $L$ and the corresponding receptor FGFRIIb ($R$):

$$\frac{\partial R}{\partial t} = \Delta R + \gamma(a - R + R^2 L) \quad \text{on } \Gamma_\Omega \quad (1)$$

$$\frac{\partial L}{\partial t} = D\,\Delta L + \gamma b \quad \text{in } \Omega \quad (2)$$

$$D\vec{n} \cdot \nabla L = -\gamma R^2 L \quad \text{on } \Gamma_\Omega \quad (3)$$

The receptor $R$ is restricted to the epithelial boundary $\Gamma_\Omega$. It is expressed with the constant rate $a$ and undergoes a linear decay $-R$. The ligand $L$ diffuses in the whole mesenchyme $\Omega_{\text{mes}}$ and is produced with the constant rate $b$. Ligand-receptor binding on the epithelial boundary results in a ligand removal at rate $-R^2L$. The quadratic term arises because the ligand dimer binds simultaneously two receptors [1], [15]. At the same time, the non-linear reaction term $+R^2L$ stimulates the receptor accumulation on the epithelial surface (which is indeed observed [16], [17]). The parameter $D$ refers to the relative diffusion coefficient of ligand and receptor, and $\gamma$ is a scaling parameter.

The outgrowth of new emerging branches is then proportional to the resulting $R^2L$-spots on the epithelial surface:

$$\vec{v}_{\text{growth}} = s \cdot R^2 L \cdot \vec{n}, \quad (4)$$

where $s$ is a growth speed scaling factor and $\vec{n}$ the normal vector on the epithelium.

Solving Eqs. (1) to (3) on image-based embryonic lung geometries shows good agreement between the numerical solution and the observed growth field [1], [7].

## Comparing the ALE and the Phase-Field Simulations

We have described the use of ALE in the COMSOL conference proceedings [2]–[4] and the Phase-Field based approach without mesenchymal growth in [5]. We start by summarising the Phase-Field implementation and extending it to include the mesenchyme as a second interface. At the same time, we restrict the ligand production to the mesenchymal border.

So far, we tracked the epithelial surface only by a phase field interface defined by $\phi_{\text{epi}} = 0$ within the epithelium and $\phi_{\text{epi}} = 1$ in the enclosing domain (see Fig. 1). To include the mesenchyme as an additional interface, we introduce another independent phase field $\phi_{\text{mes}}$ which is set to $\phi_{\text{mes}} = 0$ within the mesenchyme (and epithelium) and $\phi_{\text{mes}} = 1$ in the enclosing box $\Omega_{\text{bounding}}$ only. Thus, the mesenchymal border is defined by the interface in-between the two homogenous phases and is denoted by $\Gamma_{\text{mes}}$. Following the idea behind the phase field approach –extending the simulation domain $\Omega$ and multiplying the system of equations by the approximated characteristic function $\phi$, restricting them to the original geometries– we extend the simulation domain to $\Omega = \Omega_{\text{epi}} \cup \Omega_{\text{mes}} \cup$

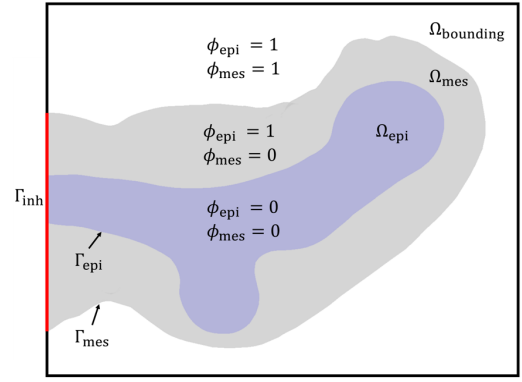

**Figure 1**: 2D Schematic of the Lung Geometry: The epithelial bulk $\Omega_{\text{epi}}$ (blue) is enclosed by the mesenchymal bulk $\Omega_{\text{mes}}$ (grey). The receptor $R$ is expressed on the epithelial-mesenchymal border $\Gamma_{\text{epi}}$ only, whereas the ligand $L$ is produced in the whole mesenchyme. In the extended model, the whole geometry is enclosed in a bounding-box $\Omega$. The mesenchymal bulk is represented by a second independent phase field equation denoted as $\phi_{\text{mes}}$ where $\phi_{\text{mes}} = 0$ within the epithelium and the mesenchyme and $\phi_{\text{mes}} = 1$ outside. The original phase-field $\phi_{\text{epi}}$ is extended to the whole domain.

$\Omega_{\text{bounding}}$ and rewrite the Turing mechanism, see Eqs. (1) to (3), as

$$\delta_{\text{epi}} \frac{\partial R}{\partial t} = \nabla \cdot (\delta_{\text{epi}} \nabla R) + \delta_{\text{epi}} \gamma (a - R + R^2 L), \quad \text{in } \Omega \quad (5)$$

$$\phi_L \frac{\partial L}{\partial t} = D \nabla \cdot (\phi_L \nabla L) + \gamma (\phi_L b - \delta_{\text{epi}} R^2 L), \quad \text{in } \Omega \quad (6)$$

where $\phi_L = \phi_{\text{epi}} - \phi_{\text{mes}}$. Thus, $\phi_L$ equals to 1 in the mesenchymal bulk only (where the ligand $L$ exists). To restrict the ligand expression to the mesenchymal surface, we multiply the production term with the approximated Dirac $\delta$-function similarly to the restriction of $R$ to the epithelial surface. Thus, Eq. (6) becomes

$$\phi_L \frac{\partial L}{\partial t} = D\nabla \cdot (\phi_L \nabla L) + \gamma (\delta_{\text{mes}} b - \delta_{\text{epi}} R^2 L) \quad (7)$$

The stabilisation terms described in [5] are added again. Moreover, to reproduce the observed *in vivo* patterning, we limit ligand production by an inhibitor that diffuses into the mesenchymal bulk from the boundary $\Gamma_{\text{inh}}$, see Fig. 1. The temporal evolution of this inhibitor $I$ is described by

$$\frac{\partial I}{\partial t} = D \, \Delta I - k_d I$$
$$\frac{\partial I}{\partial t} = p_0$$

where $k_d$ is the degradation rate and $p_0$ the constant inhibitor inflow from the stalk. In the Phase-Field description, the above system of equations reads as

$$\phi_L \frac{\partial I}{\partial t} = D \nabla \cdot (\phi_L \nabla I) - k_d \phi_L I \qquad (8)$$

Finally, the production of $L$ is damped by $f(I) = (1 + I^2)^{-1}$ resulting in

$$\phi_L \frac{\partial L}{\partial t} = D \nabla \cdot (\phi_L \nabla L) + \gamma \big(\delta_{\mathrm{mes}} f(I) b - \delta_{\mathrm{epi}} R^2 L\big) \qquad (9)$$

Thus, we solve Eqs. (5) and (9) for the receptor-ligand Turing pattern as well as (8) for reproducing the ALE simulations.

## Use of COMSOL Multiphysics® Software

We use COMSOL Multiphysics® 5.3 to perform all the simulations and ParaView for the post-processing tasks. A more detailed description of the Phase Field implementation in COMSOL Multiphysics® can be found in [5], the details of the ALE-implementation in [2]–[4].

## Results

We first study the behaviour of the *Parameter controlling the interface thickness $\varepsilon$* of the Level-Set module on the effective interface thickness. The finer this interface, the better we approximate the explicit surface of the geometry. On the other hand, the effective interface thickness implies an upper limit on the mesh resolution and thus maximum mesh size in this region. The follow-up paper of the referenced phase field method for the Level-Set module [18] states the relation $d \approx 6\varepsilon$, where $d$ is the effective interface thickness [19]. Fig. 2 shows the effective interface thickness defined by $0.05 \leq \phi \leq 0.95$ depending on different $\varepsilon$ values. The blue line is the linear regression and confirms the relation $d \approx 6\varepsilon$.

The next important aspect is the conservation of the area under the approximated Dirac $\delta$-function in the direction of the interface normal:

$$A_\delta = \int_L \delta \, ds = \int_L |\nabla \phi| ds \approx 1.$$

If this does not hold due to numerical errors, the reaction and the production rate on the surface is not comparable to the ALE-setting. Fig. 3 shows the derived line integral by COMSOL over $\delta \approx |\nabla \phi|$ in 3D. The area under the $\delta$-function is correctly obtained even for a fine interface (and thus a steep and narrow approximated Dirac $\delta$-function).

Next, we compare the convergence rate of the two methods and determine the maximal interface thickness of the Phase-Field implementation to get accurate results. We do this in the static case and thus without the need of the Moving Mesh module. Nevertheless, we call this implementation ALE as it resembles the ALE implementation in the growing case. As we do not have an analytic solution, we compare the solutions to the solution on the finest mesh ($h_{\max} = 6\varepsilon/5$ on the epithelial surface with $\varepsilon = 0.0625$). The resulting Turing pattern can be seen in Fig. 4A. In Fig. 5 we show the convergence behaviour of the ALE implementation for $\varepsilon \in \{0.125, 1, 2, 4, 8, 16\}$ on the epithelial surface in the $L^2$- and $L^\infty$-norm. The solution converges in between $O(h^{3/2})$ and $O(h)$. As we evaluate the volumetric solution on the surfaces only, we expect a loss of order $1/2$ in the convergence due to the trace theorem. As we do not have nested meshes (and no influence on the iso-surface discretisation), we have to interpolate the solution $u_h$ on the finer mesh on the coarser. We map the solution from the finer mesh to the coarser one as follows: we identify each node in the coarser mesh with the nearest node in the finer mesh. Now, on the coarser mesh, we can approximate the relative $L^\infty$-norm and the $L^2$-norm. Given the continuity of the solution, the interpolation error by the mapping depends on the maximum distance between the vertices of the two meshes. The relative distance is in the order of $10^{-4}$ and thus do not alter the approximated relative errors significantly. For $\varepsilon \in \{0.25, 0.5\}$, we get an additional streak between two Turing spots resulting in a higher relative error.

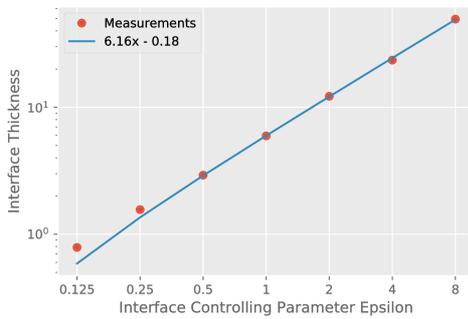

**Figure 2: Effective Interface Thickness:** The linear regression fit confirms the relation $d \approx 6\varepsilon$. The interface thickness is defined as $0.05 \leq \phi \leq 0.95$.

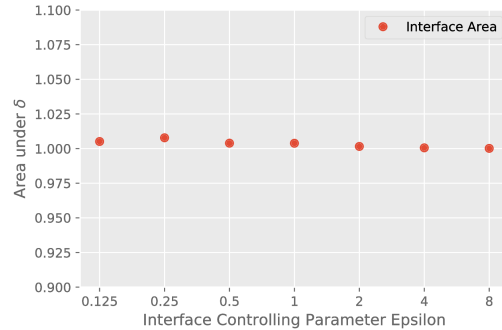

**Figure 3: Area under δ:** The result of the numerical line integration over $\delta \approx |\nabla \phi|$ is shown. COMSOL manages to evaluate it independently of the mesh size.

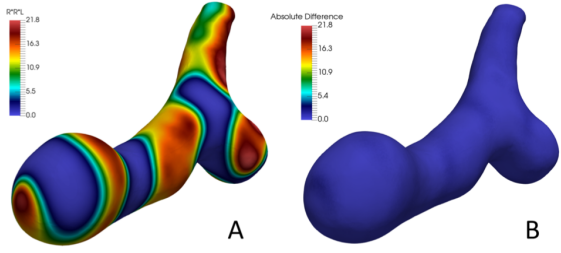

**Figure 4: Turing Pattern of the Finest Solution and Difference of the Phase-Field Solution to the ALE Solution:** (A) Turing Pattern for the parameters $\gamma = 0.1$, $a = 0.03$ and $b = 0.25$ solved on the finest mesh with $\varepsilon = 0.0625$ (with the ALE implementation). (B) The absolute difference of the Phase-Field solution with $P2$ elements (phase-field equations only) and a mesh size of $\varepsilon = 0.5$. Same Turing parameters as in the ALE simulation.

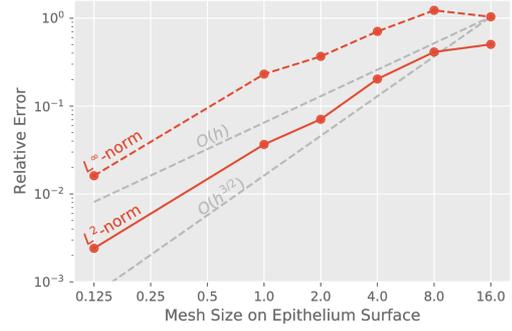

**Figure 5: Convergence Analysis for the ALE Solution:** The mesh size on the epithelium is set to $\varepsilon \in \{0.125, 1, 2, 4, 8, 16\}$ and the solution with the finest mesh ($\varepsilon = 0.0625$) is taken as the reference solution. The convergence order is between $O(h)$ and $O(h^{3/2})$. The latter is the theoretically maximal achievable rate taking into account the trace theorem for evaluating on a surface a columetric solution.

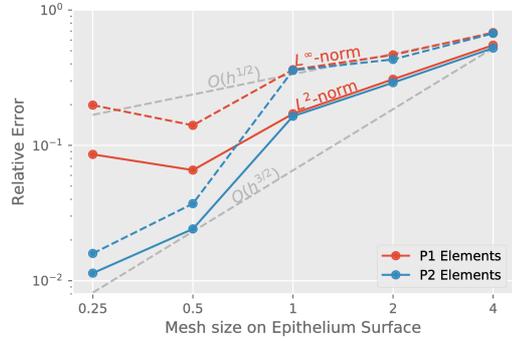

**Figure 6: Convergence Analysis for the Phase-Field Simulation:** The relative error compared to the ALE solution with $\varepsilon = 0.125$. Mesh sizes vary by $\varepsilon \in \{0.25, 0.5, 1, 2, 4, 8\}$. We plot the relative $L^2$- and $L^\infty$-error (normal and dashed line) for $P1$ elements (red) and $P2$ elements in blue.

Now, we compare the solution of the Phase Field simulation to the ALE solution. Fig. 6 shows again the convergence for $h_{max} = 6\varepsilon/5$ where $\varepsilon \in \{0.25, 0.5, 1, 2, 4\}$. This time, we have to mesh the geometry within the epithelium, too. Using this relation for $h_{max}$, we can ensure that we have at least 5 elements in the interface and thus resolving the epithelium surface good enough. The convergence order depends on the element order as well as the maximal mesh size $h_{max}$. The simulations with $P1$ elements converge only with a rate of $O(h)$ as plotted in red. The solution with the $P2$ mesh elements converges similarly for $\varepsilon \geq 1$. However, for the configuration with $\varepsilon < 1$, we reach a convergence order of up to $O(h^2)$. Thus, $P2$ elements and a maximum $\varepsilon < 1$ is needed for the Level-Set module to get accurate and to the ALE implementation comparable result. Smaller mesh sizes were computationally not possible as we already get around 4 million DoF to solve for $\varepsilon = 0.5$ (compared to 1 million DoF in the ALE version). The same setting in the ALE-implementation is also 11 times faster (for the static case).

The absolute difference of the Phase Field solution with $P2$ elements and $\varepsilon = 0.5$ is shown in Fig. 4B. For the growth simulations, we had to use $P1$ elements and $\varepsilon = 1$ due to the computational burden.

For the growth simulations, we extended the model such that the mesenchymal border is represented as a second phase field interface as above. We set $\varepsilon = 4$ for this mesenchymal Level-Set module and $P1$ elements as there are no reactions happening on this surface and thus does not be as thin and fine-meshed. The ALE solution at $t = 250s$ is shown in Fig. 7A. For this setting, the maximum possible simulation time is around $t = 270s$ after which the re-meshing step fails due to intersecting edge elements.

The Phase-Field based simulation results are shown in Fig. 7B-7D. The change of the mesenchymal surface from $t = 250s$ to $t = 500s$ and $t = 500s$ to $t = 1000s$ is indicated by the darker contour. The growing branches veer off into the free available space within the mesenchyme and do not penetrate the (implicit) mesenchymal surface. However, new emerging branches still grow together if there is not enough space to elude, see [5]. We observe this phenomenon more often with bigger $\varepsilon$ and the resulting mesh sizes. A faster mesenchymal growth speed was not possible so far as this would need a finer mesh around the mesenchyme. Compared to the ALE-simulation, the Phase-Field based simulations are more stable and do not crash as we do not have to re-mesh or deform the mesh. At the same time, we get a similar branching behaviour as in the ALE-model but with more nodular branch tips. Thus, our surface normals in the Phase-Field simulations differ compared to the ALE-model und thus influence the outgrowth direction differently. Comparing to image-based data, the emerging buds are more nodular, too (data not shown). Another important observation is that the effective growth speed of the ALE simulations is not the same as for the Phase-Field simulations. This is mainly because we have to restrict the velocity field to the surface more strictly by multiplying with the (not normalised) Dirac $\delta$-function. We have to do this for numerical stability reasons, and further analysis needs to be done. The maximum value of the approximation of $\delta$ is also dependent on the interface thickness based on $\varepsilon$ and thus rather difficult to fine-tune. Another factor is the Reinitialization parameter $\gamma_{LS}$ which controls the dynamic of the underlying phase-field interface. If the value is set too large, the newly emerged branches are smoothed out and disappear again. If the value is set too small, the interface starts to oscillate and introduces numerical errors in the form of small artificial bulks (data not shown). At the same time, this parameter controls the rigidity of the phase-field interface and thus influences the outgrowth speed.

## Conclusions

Our Phase-Field approach for simulating embryonic branching morphogenesis reproduces the static solution of the current ALE-based simulations down to a relative error < 2%. The dynamic growth simulation is more stable than the ALE simulations as in the Phase-Field approach we do not have to distort the underlying mesh at all. However, this comes at the cost of computational more demanding simulations. The mesh has to resolve the phase field interface in the normal direction well enough. At the same time, the interface thickness needs to be small enough to get to this relative error below 2%, transforming the simulation into a spatial multi-scale problem. Additionally, we have to mesh the full epithelial geometry and not just its surface as in the ALE model. This results in an 11 times longer run time compared to the ALE simulation with a mesh of the same epithelial resolution (in the static case). Similarly, Aland et al. show comparable results (24 times slower) for the Taylor flow problem solved with sharp and diffuse interfaces [20]. This problem is amplified in the growth simulations with the extended bounding box and the mesenchymal surface as an additional phase-field interface. Thus, the ALE approach should be favoured for prototyping and parameter screens in the static setting. For the growth scenario (e.g. based on the optimal configuration from a parameter screen with the ALE solution), the Phase-Field approach leads to a more stable simulation.

Increasing the interface thickness exacerbates another drawback of the implicit boundary methods. Eventhough the phases are strictly separated by the interface, two branches can fuse together if they come close enough. There exists some early work on topology-conserving Phase-Field methods where the merging is eliminated by a penalisation term based on the geodesic distance between two points. However, this approach needs to couple the FEM solver with Dijkstra's algorithm to find the shortest path between two points [21] and is thus not easy to implement in COMSOL.

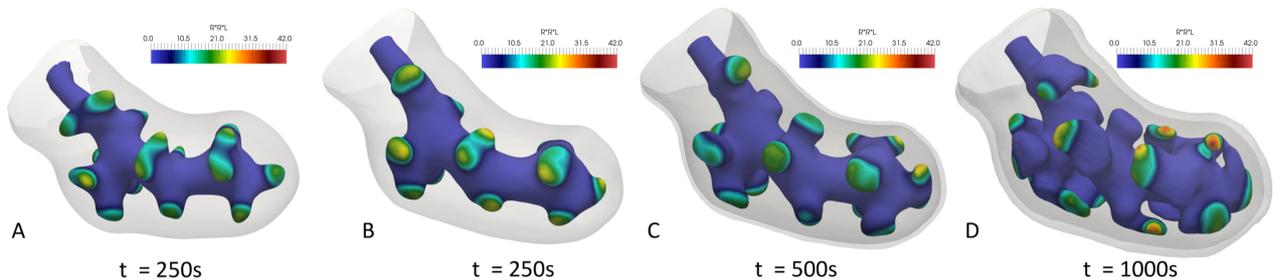

**Figure 7: Growth Simulations: (A)** shows the ALE-based simulation at $t = 250s$. Around $t = 270s$ the simulation fails due to re-meshing problems. **(B – D)** show the Phase-Field-based simulations for the same parameter configuration as for the ALE simulation for $t \in \{250s, 500s, 1000s\}$.

Nevertheless, this branch fusing behaviour is observed in the embryonic pancreas morphogenesis and thus a welcome feature for simulating pancreas growth. For the embryonic lung simulations, the only way to circumvent this behaviour is to use a smaller interface thickness.

The plug & play-nature of COMSOL allows us to easily implement the modifications to the receptor-ligand based Turing mechanism and to link it to one of the existing phase field modules. The time-dependent solver manages to detect the non-linear Turing instabilities out-of-the-box, such that first Turing patterns are reached rather easily. However, there are some desirable features missing. The built-in adaptive mesh refinement is not optimal for our use case. First, it is not possible in COMSOL to enforce a certain number of elements within the interface (even-though the follow-up paper of the Level-Set module implementation explains a way to achieve this [19]). Second, the available settings are designed for moving domains and not time-dependent growing domains. As our approach introduces a CFL-like constraint on the maximum time step depending on the current velocity field, we have to limit the maximum time step of the BDF-solver adaptively such that the interface does not move further than half of the interface thickness. Otherwise, we would lose the constantly extended concentration of $R$ (see [5]) in the interface. However, this is currently not possible in COMSOL as only global parameters and not derived values are accepted. Thus, we have to approximate the maximum possible value manually in advance.

## Acknowledgements

We thank our colleagues for discussion and the COMSOL support staff for their excellent support, especially Sven Friedel, Zoran Vidakovic and Thierry Luthy.